\documentclass[conference]{IEEEtran}
\IEEEoverridecommandlockouts
\usepackage{cite}
\usepackage{amsmath,amssymb,amsfonts}
\usepackage{graphicx}
\usepackage{textcomp}
\usepackage{xcolor}
\usepackage{algorithm}
\usepackage{algpseudocode}
\usepackage{subfigure}
\usepackage{multirow}
\def\BibTeX{{\rm B\kern-.05em{\sc i\kern-.025em b}\kern-.08em
    T\kern-.1667em\lower.7ex\hbox{E}\kern-.125emX}}

\begin{document}

\title{MAC Address De-Randomization using Multi-Channel Sniffers and Two-Stage Clustering}

\author{\IEEEauthorblockN{Giovanni Baccichet, Corrado Innamorati, Alessandro E. C. Redondi, Matteo Cesana}
\IEEEauthorblockA{\textit{Dipartimento di Elettronica, Informazione e Bioingegneria, Politecnico di Milano} \\
email: \{name.surname\}@polimi.it}
}

\maketitle

\begin{abstract}
MAC randomization is a widely used technique implemented on most modern smartphones to protect user's privacy against tracking based on Probe Request frames capture. However, there exist weaknesses in such a methodology which may still expose distinctive information, allowing to track the device generating the Probe Requests. Such techniques, known as MAC de-randomization algorithms, generally exploit Information Elements (IEs) contained in the Probe Requests and use clustering methodologies to group together frames belonging to the same device. While effective on heterogeneous device types, such techniques are not able to differentiate among devices of identical type and running the same Operating System (OS). In this paper, we propose a MAC de-randomization technique able to overcome such a weakness. First, we propose a new dataset of Probe Requests captured from devices sharing the same characteristics. Secondly, we observe that the time-frequency pattern of Probe Request emission is unique among devices and can therefore be used as a discriminative feature. We embed such a feature in a two-stage clustering methodology and show through experiments its effectiveness compared to state-of-the-art techniques based solely on IEs fingerprinting. The original dataset used in this work is made publicly available for reproducible research. \\
\end{abstract}


\section{Introduction}
\label{sec:introduction}
With millions of Wi-Fi-enabled devices being shipped every year, we are truly living in a \textit{Global Village}, where every individual becomes more and more connected. The ubiquity of connected devices in our daily lives offers numerous advantages, but it also raises concerns about personal privacy. 
Indeed, Wi-Fi-enabled devices repeatedly transmit specific unencrypted signals, known as Probe Request frames, to scan the surrounding area, searching for nearby networks. Like every other Wi-Fi frame, Probe Requests contain a unique device identifier, the source Media Access Control (MAC) Address, a fixed and exclusive piece of information linking to personal information. A passive attacker may therefore collect Probe Request frames,
hence obtaining sensitive information regarding the device’s owner. 
The rampant diffusion of Wi-Fi devices and an evolving concern for privacy have led to a search for securing
techniques, to prevent antagonists from tracking users’ persistent identifiers. So far, one of the most popular countermeasures consists of MAC Address Randomization, where a device generates randomized addresses to be used in a pseudo-random pattern during the Probe Request emission process.
MAC address randomization is not a standardized technique: consequently, the implementation of privacy-preserving schemes varies depending on the OS and vendors implementing this practice \cite{fenske2021three}. Even though randomization has observed a general improvement and
deployment diffusion in recent years, multiple logic flaws and residual information can
still threaten users’ privacy and lead to possible tracking vulnerabilities. Overcoming randomization is specifically a challenge for a variety of mobility-based application fields, including those linked to urban planning, citizen safety, the deployment of public transportation services, and the design of telecommunications services.
Many works in the literature have studied MAC randomization issues and limitations, evaluating possible techniques and methodologies to defeat randomization or showing how it can still be
bypassed to get specific device-related information.
Such works generally rely on IEs contained in the Probe Requests as features to perform MAC address de-randomization through fingerprinting. However, such features are identical across different devices sharing the same characteristics (e.g., two iPhones 12 running the same version of iOS) and are therefore useless for MAC de-randomization in the presence of many devices of the same type, which often happens in reality.
In this paper, we target specifically this scenario and propose a novel MAC address de-randomization technique through the following contributions:
\begin{itemize}
\item \textit{Dataset augmentation:} we collect Probe Requests from several modern mobile phones following the same procedure described in \cite{Pintor2022-dataset}. In detail, we focus on creating a much more challenging dataset, capturing Probe Request frames from devices sharing the very same characteristics in terms of model type and version, OS, etc. The dataset is made publicly available for reproducible research.
\item \textit{Multichannel features:} we design new features for performing MAC de-randomization, by exploiting the peculiar probing behavior of each device across different frequency channels.
\item \textit{Two-stage clustering:} we propose a new strategy based on two-stage clustering for performing MAC de-randomization, showing its performance with extensive tests performed over the new collected dataset, as well as the dataset provided in \cite{Pintor2022-dataset}.
\end{itemize}
The paper is organized as follows: Section \ref{sec:related_work} discusses prior research on Probe Requests and MAC de-randomization. Section \ref{sec:dataset} details the new acquired dataset, while Section \ref{sec:methodology} outlines the proposed methodology. Section \ref{sec:results} discusses experimental results, while Section \ref{sec:conclusion} concludes the paper.





\section{Related work}\label{sec:related_work}
The theme of randomization and its faults has interested many studies since its first introduction in the art. Early studies were focused on the effectiveness of this new technique concerning the tracking of mobile
devices.
In 2016, Vanhoef et al. \cite{vanhoef2016mac} analyzed the limits of randomization showing how multiple
features and IEs from Probe Request frames could still be used to track
mobile devices. The authors also identified possible attacks to defeat randomization and directly recover the global MAC address of a device. Such studies were based on three
major datasets: the \textit{Lab}, \textit{Sapienza} and \textit{Station} datasets, each containing hundreds of captured frames in specific locations and collected around 2013-2015 \cite{barbera2013signals}. IEs extracted from Probe Requests were analyzed in terms of information impact and entropy, demonstrating encouraging results in terms of fingerprinting. 
Matte et al. \cite{matte2016defeating} proposed a novel strategy based on timing observation of the delivered Probe Request frames, building an Inter-Frame Arrival Time signature to recognize and
track devices, with a success rate of approximately 77\%.
In 2017, Martin et al. \cite{martin2017study} observed how randomization implementation on mobile devices was sporadic and variable between manufacturers, also failing to hide the global MAC
address in certain conditions.
Some observations can be made regarding this context. First, the aforementioned studies were performed in a relatively "early" diffusion of randomization. The used datasets
date back to 2013 when randomization did not exist. Moreover, the collected data was
anonymized, giving no ground truth over which testing the method results. Since then,
manufacturers have more thoroughly adopted randomization, which saw a more defined
implementation. Studies in \cite{martin2017study} demonstrated how in more recent datasets (2017), randomization saw a rising adoption concerning the ones previously mentioned. In \cite{fenske2021three} (2021) a more up-to-date analysis is provided, confirming the referred trend. Authors have performed a cross-sectional study over 160 devices, to test randomization implementation, schemes, or eventual tracking vulnerabilities. The efficacy of the techniques
introduced in previous works was tested, to observe whether the randomization approach
has changed and if faults were still an issue. Results showed how updated OSes assured a high privacy degree, leading to a more standard randomization application. Some of the de-randomization and tracking methodologies illustrated in previous studies demonstrated a lack of efficacy, showing how randomization is evolving, limiting
possible traces or remnants of sensitive information to be exposed. However, many mobile devices were still failing to prevent tracking, especially the ones supporting older OS
versions. Authors have analyzed how the lack of a standardized randomization definition
has led to ``fragmentation and significant differences in effective privacy from one device
to another".
Recent research considered all of the aforementioned remarks and focused on improved de-randomization techniques with innovative approaches, mostly focusing on the IEs content,
which has been observed to be device dependent, and specific association methodologies.
Uras et al. \cite{uras2020wifi} introduced a method to overcome the randomization approaches in tracking and smart cities scenarios. The suggested technique uses clustering algorithms over IEs features extracted from Probe Requests, specifically DBSCAN, and OPTICS, to determine the number of devices, showing encouraging final results. The dataset used in this work is detailed in \cite{Pintor2022-dataset} and constitutes one of the few examples of labeled datasets based on Probe Requests. The dataset contains labeled Probe Requests from 22 known devices, captured in isolated conditions for 20 minutes over three Wi-Fi channels (1,6 and 11). The availability of ground truth data allows to compute precise performance metrics in terms of MAC de-randomization and device counting capabilities, as done in 
\cite{pintor2022analysis} and
\cite{uras2022mac}.
We highlight that the dataset in \cite{Pintor2022-dataset} is very heterogeneous, with just two devices sharing the same manufacturer, model, and OS. This constitutes an optimistic scenario, as different types of devices are naturally characterized by differences in the IEs contained in the emitted Probe Requests (due to the different OS versions, Wi-Fi chipset, and other internal characteristics). The goal of this work is therefore twofold: first, to provide a more challenging dataset, characterized by devices of the same manufacturer, type/version, and running the same OS. Second, propose a new approach for MAC de-randomization and test its performance on the newly provided dataset.

\begin{table}[t]
\caption{Dataset Expansion with device model, software version, and identifier}
\label{tab:new-devices}
\begin{center}
\begin{tabular}{|c|c|c|}
\hline
\textbf{Device} & \textbf{OS}& \textbf{Identifier} \\
\hline
Apple iPhone 7& iOS \texttt{15.5}& iPhone7-F \\
\hline
Apple iPhone 11& iOS \texttt{16.4.1(a)}& iPhone11-B \\
\hline
Apple iPhone 11& iOS \texttt{16.4.1(a)}& iPhone11-F \\
\hline
Apple iPhone 11& iOS \texttt{16.4.1(a)}& iPhone11-M \\
\hline
Apple iPhone 11& iOS \texttt{16.4.1(a)}& iPhone11-C \\
\hline
Apple iPhone XR& iOS \texttt{16.4.1(a)}& iPhoneXR-A \\
\hline
Apple iPhone XR& iOS \texttt{16.4.1(a)}& iPhoneXR-L \\
\hline
Apple iPhone 12 & iOS \texttt{16.4.1(a)}& iPhone12-M \\
\hline
Apple iPhone 12 Pro& iOS \texttt{16.4.1(a)}& iPhone12Pro-C \\
\hline
Samsung S21 Ultra& Android \texttt{13}& S21Ultra-M \\
\hline
Oppo Find X3 Neo& Android \texttt{13}& OppoFindX3Neo-A \\
\hline
\end{tabular}
\end{center}
\end{table}

\section{Dataset}
\label{sec:dataset}

\subsection{Capturing Probe Requests}
To provide a more challenging scenario, we consider 11 new devices as reported in Table \ref{tab:new-devices}. 
As one can see, there is large homogeneity among the chosen devices, with many repeated models (e.g., iPhones 11) as well as models already present in the dataset in \cite{Pintor2022-dataset} (e.g., iPhone 7 and 12). In addition, we favor Apple devices over Android-based ones, as the latter are the vast majority in \cite{Pintor2022-dataset}. 

For the sake of consistency, we carefully follow the procedure detailed in \cite{Pintor2022-dataset} for the construction of the Probe Request dataset. In detail, we place each device in a controlled and isolated environment. In particular, we captured traffic using three ALFA AWUS036ACS Wi-Fi interfaces based on the Realtek \texttt{RTL8812AU} chipset, which is known for excellent performance in traffic monitoring and sniffing tasks. The three interfaces are tuned to Wi-Fi channels 1, 6, and 11 in the 2.4 GHz band, respectively, and are programmed to capture Probe Request frames independently and simultaneously. During each capture, the device under consideration is operated in different modes (e.g., screen on/off, power saving on/off, etc.) as reported in \cite{Pintor2022-dataset}. Once the capture is completed we add to the obtained dataset data from  \cite{Pintor2022-dataset} and obtain two new datasets, named \textit{Twin Companions} and \textit{Complete} dataset, respectively. In the \textit{Twin Companions} dataset we add to our capture data from similar devices in \cite{Pintor2022-dataset}, to maximize device homogeneity. Namely, we add to the 9 newly captured devices having a twin (that is, not considering the Samsung S21 and the Oppo Find X3) the two Google Pixel 3A and the iPhone 7, 12 and XR contained in \cite{Pintor2022-dataset}, for a total of 14 devices.
Conversely, in the \textit{Complete} dataset, we add all data from \cite{Pintor2022-dataset} to our capture, resulting in a total of 33 devices, 14 of which sharing
the same model and software version as at least one other
device. 

\subsection{Data pre-processing}
We pre-process the captured Probe Request frames closely following what was employed by Pintor et al. In detail, we process the raw captured frames with the \texttt{scapy} Python library and extract IEs from each captured frame. The following IEs are considered: HT Capabilities (\texttt{ID = 45}), Extended Capabilities (\texttt{ID = 127}), and Vendor-Specific Tags (\texttt{ID = 221}). When a field was empty or absent, its value was set to 0, while numeric fields retained their numeric values. Arrays were transformed into the sum of their values, and strings were converted by summing the ASCII values of each character. 

Additionally, we extracted the  Channel and DS Parameter Set (in particular, the Current Channel field, hereafter referred to as DS Channel). This addition enables analysis of multi-channel behavior in Probe Request bursts.

Finally, we extract the source MAC address and the capture timestamp of each Probe Request frame. We concatenate all the extracted information in a feature vector and we label it with the ground truth MAC address of the emitting device.

Notably, five devices in the \textit{Complete} dataset did not implement MAC address
randomization (i.e., the source MAC address contained in the Probe Request frames is identical to the ground truth MAC address). However, they were still included in the analysis as
the source MAC address is not used as a feature in the following.

\begin{figure}[t]
    \centering
    \includegraphics[width=1\linewidth]{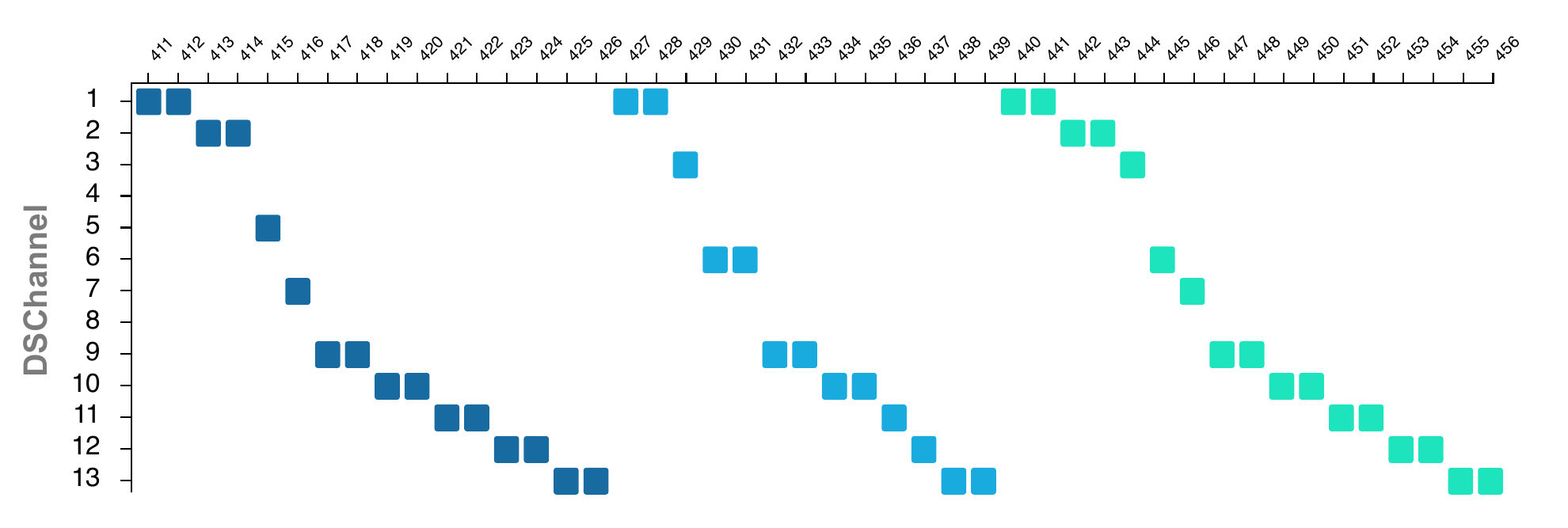}
    \caption{Google Pixel 3A (labeled L) DS Channel Pattern Over Time, MAC Addresses are distinguished by color.}
    \label{fig:pixelL-pattern}
\end{figure}

\begin{figure}[t]
    \centering
    \includegraphics[width=1\linewidth]{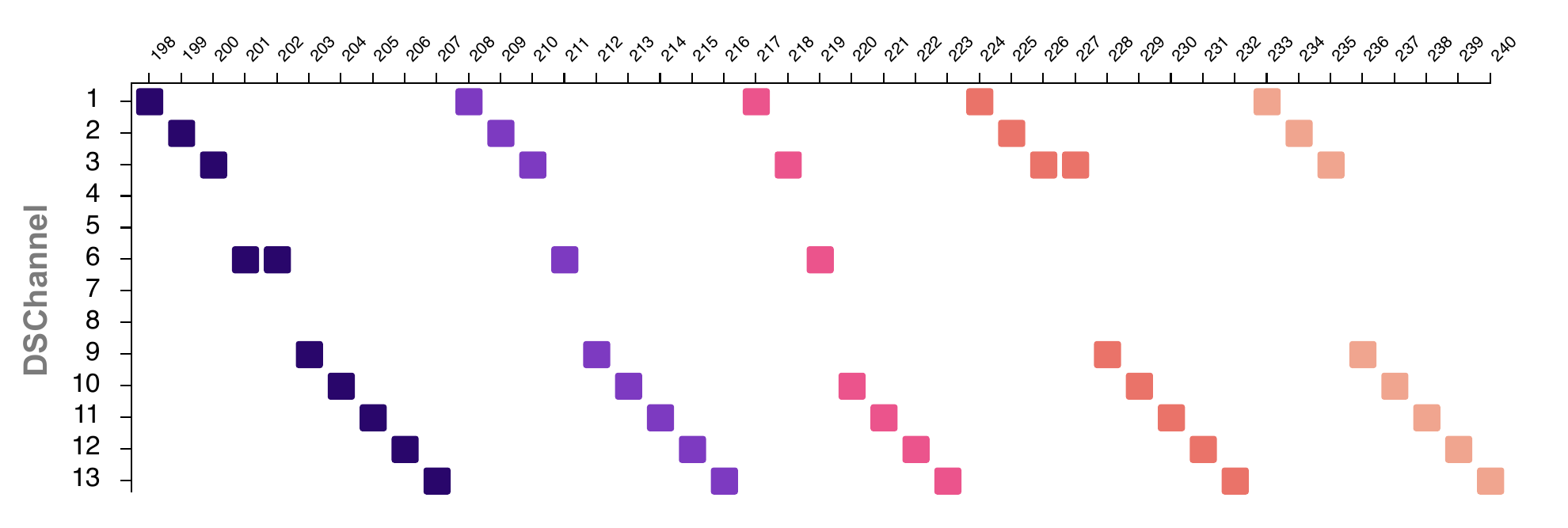}
    \caption{Google Pixel 3A (labeled V) DS Channel Pattern Over Time, MAC Addresses are distinguished by color.}
    \label{fig:pixelV-pattern}
\end{figure}

\subsection{Multi-channel feature}

Probe Requests are generally transmitted in bursts sweeping all Wi-Fi channels, to maximize the likelihood of finding a replying Access Point (AP). When MAC randomization is used, bogus source MAC addresses are continuously produced and used by the emitting devices. Although the specific details are device-dependent, an observed general trend is to change the randomized MAC address at each Probe Request burst.

While previous studies have extensively examined content-based and timing-based features for MAC de-randomization, channel-based attributes have received relatively less attention. We designed our approach to fill this gap by conducting a comprehensive multi-channel behavior analysis. We specifically focus on monitoring the Channel and DS Channel fields in Probe Requests that share the same MAC address and, by extension, belong to the same burst of requests.

Figures \ref{fig:pixelL-pattern} and \ref{fig:pixelV-pattern} show the time-frequency Probe Request emission pattern of the only two identical devices (Google Pixel 3A) of the dataset in \cite{Pintor2022-dataset}. The horizontal axis represents the packet number (time), while the vertical axis depicts the DS Channel field (frequency). Colors represent bursts of Probe Requests with the same source MAC address. 
As one can see, the time-frequency behavior of the probing pattern is different although the two devices are identical. A possible explanation could reside in the Preferred Network List (PNL) existing on each device, which contains information on already visited Wi-Fi networks, such as SSIDs and channels. PNLs are generally used to drive the Probe Requests emission process and search for known networks first. Therefore, identical devices with different PNLs are characterized by different probing patterns. 

We encode such observation in a feature vector that could be efficiently employed, by first grouping together all Probe Requests sharing the same source MAC address (color), and then describing the burst with a vector that sequentially records the DS Channel based on the order of arrival of each frame. Let $f_i$ be the DS Channel observed in the $i-th$ captured probe request of a burst, with $i = 1\ldots L$ and $L$ the burst length. We concatenate all $f_i$ in a vector $F= [ f_1, f_2, \ldots, f_L ]$.

For instance, the vector representing the first burst visible in Figure \ref{fig:pixelL-pattern} is \texttt{[1, 1, 2, 2, 5, 7, 9, 9, 10, 10, 11, 11, 12, 12, 13, 13]}. Note that bursts may have different lengths $L$: therefore, the arrival order feature vector is zero-padded to a fixed length, equal to the maximum observed one $L_{\text{max}}$, that is $F \in \mathbb{N}^{L_{\text{max}}}$.

Finally, each burst is described by concatenating the IEs features, which are stable across all frames in the burst, and the arrival order vector $F$.


\section{Clustering Methodology}
\label{sec:methodology}

The features extracted from the bursts of Probe Requests are input to a clustering algorithm with a twofold objective: (i) group together bursts having different source MAC addresses, but belonging to the same device (i.e., performing de-randomization) and (ii) separate devices of the same model/type. 

For the task at hand, we adopt a two-stage process:
\begin{itemize}
\item \textit{Coarse-grained clustering}: the first step uses solely the IEs features extracted from Probe Requests to group together bursts sharing common characteristics. Such an approach has already been used successfully in a number of previous works \cite{uras2020wifi,uras2022mac,pintor2022analysis} and is therefore selected as a preliminary step. In detail, we rely on the DBSCAN algorithm. The aim of this first step is to produce clusters whose \textit{completeness} is maximized, that is grouping together all bursts coming from the same physical device. Note that the resulting clusters may be characterized by low \textit{homogeneity}, as Probe Requests from identical devices may be grouped in the same cluster. 

\item \textit{Fine-grained clustering}: the second clustering step works independently on each cluster produced by the first stage, and has the objective of breaking down a cluster in multiple sub-clusters characterized by greater \textit{homogeneity}. 
For this task, the multi-channel feature vector $F$ is used as input to the $k$-means clustering algorithm, using cosine similarity as a distance metric in the computation of the clusters. We opt for cosine similarity rather than Euclidean distance due to the nature of the feature vector $F$, disregarding its actual magnitude. 
\end{itemize}

Both steps require tuning the hyperparameters controlling the clustering behavior: the following describes the approach used for selecting such parameters in the two stages. 

\begin{figure}[t]
    \centering
    \includegraphics[width=0.78\linewidth]{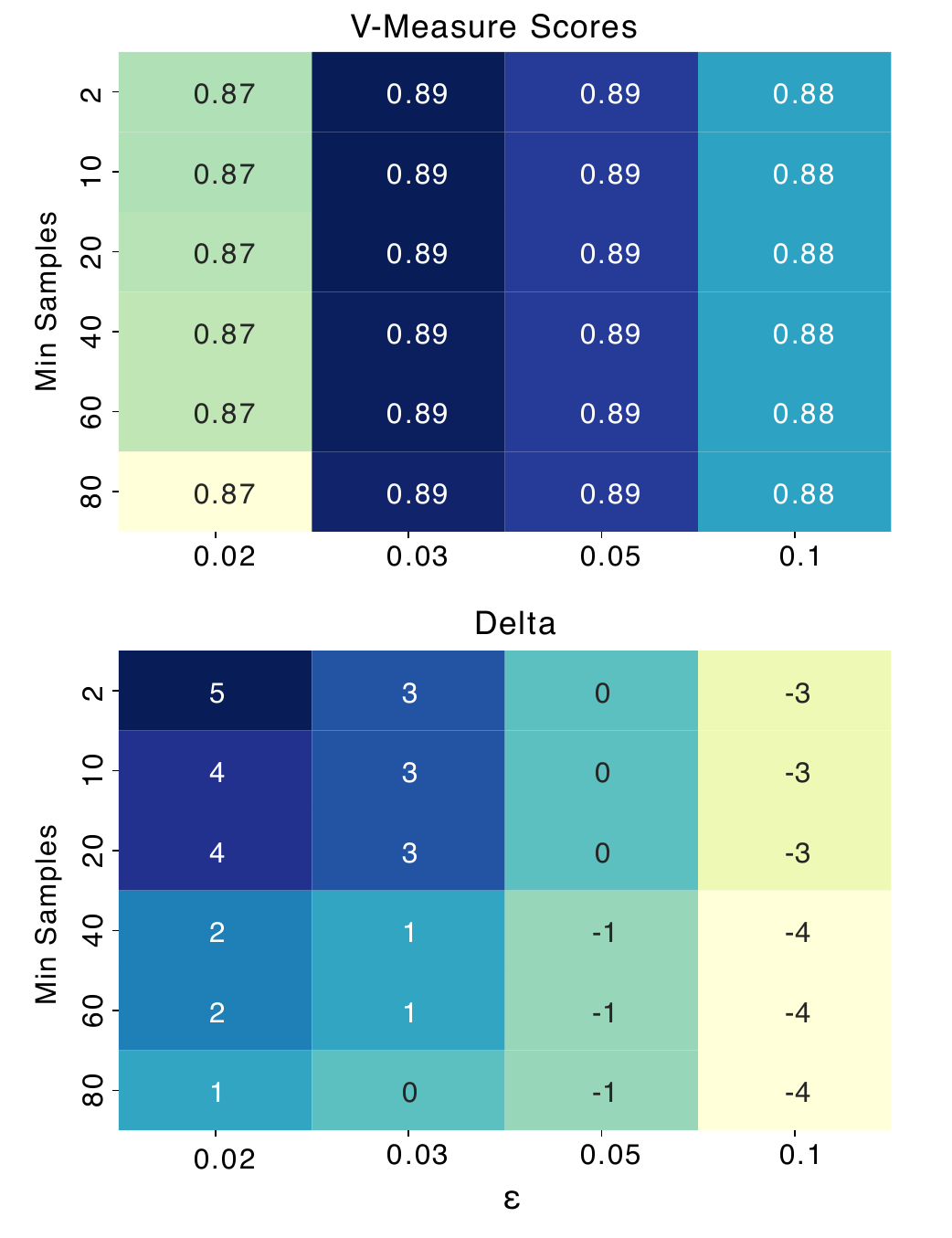}
    \caption{Delta and Cluster Homogeneity: DBSCAN Algorithm Applied to IEs for different hyper-parameters.}
    \label{fig:hyper-tuning}
\end{figure}

\subsubsection{DBSCAN parameters}
DBSCAN is a density-based clustering algorithm designed to automatically determine the number of clusters within a dataset while effectively identifying clusters of varying shapes and sizes based on the distribution of data point densities. The algorithm requires two input parameters: $\varepsilon$, defining a radius around each observation defined as $\varepsilon$-neighbourhood, and MinPts, the minimum number of samples required in the $\varepsilon$-neighbourhood to create a cluster. To train optimal values for the two parameters in the most generalized manner, we work as follows. First, we select from the original datasets (either the \textit{Twin Companions} or the \textit{Complete}) subsets of devices of different cardinality $c$. For each cardinality, multiple subsets are produced by randomly selecting $c$ devices among the ones in the original dataset. Then, we run DBSCAN over each subset with different input parameters and we compute two performance metrics, namely:
\begin{itemize}
\item V-measure: the harmonic mean between clusters homogeneity and completeness \cite{rosenberg2007v}. A cluster is homogeneous if it contains only data belonging to a single device. At the same time, a cluster is complete if it contains all data points from a device. 
\item Absolute error (Delta): the difference between the subset cardinality $c$ and the number of clusters produced by DBSCAN.
\end{itemize}
Figure \ref{fig:hyper-tuning} reports the two performance metrics, averaged over all subsets and for all parameters tested. We select as final values $\varepsilon = 0.05$ and a $MinPts = 10$, which allows us to maximize the V-measure metric while minimizing the Delta error. 

\subsubsection{$k$-means}
For the second stage, we employ $k$-means over the time-frequency feature vectors $F$, by breaking down each cluster produced by DBSCAN into multiple subclusters. To select the number of clusters $k$ we run the algorithm several times, setting $k$ from 1 to a predefined maximum value (5, in our case). At each time, we compute the within-cluster sum of squares cosine similarities with the respective cluster center as a distortion metric. As $k$ increases, the distortion decreases, making it a crucial criterion for selecting $k$. Finally, we apply the so-called elbow rule by computing the cumulative sum of the normalized derivative of the distortion with respect to $k$, selecting the optimal $k$ at a specific threshold. Rather than using a fixed threshold, we link it to the original similarity structure within each pre-cluster, computed as the average cosine similarity among all observations in the cluster. We find the following rule effective:

\begin{equation}
    \text{threshold = 0.4 + 0.6 $\cdot$ (1 - max\{0, avg(similarity)\})}
    \label{eq:dynamic-threshold}
\end{equation}

When the original cluster variance is high (low similarity), the threshold is increased, promoting more clusters. Conversely, when data points exhibit high similarity, the threshold is decreased, encouraging fewer clusters or even stopping the second clustering stage. This adaptability ensures the clustering process is data-driven, accommodating variations in the data's intrinsic structure and context.
\begin{figure}[t]
    \centering
    \includegraphics[width=1\linewidth]{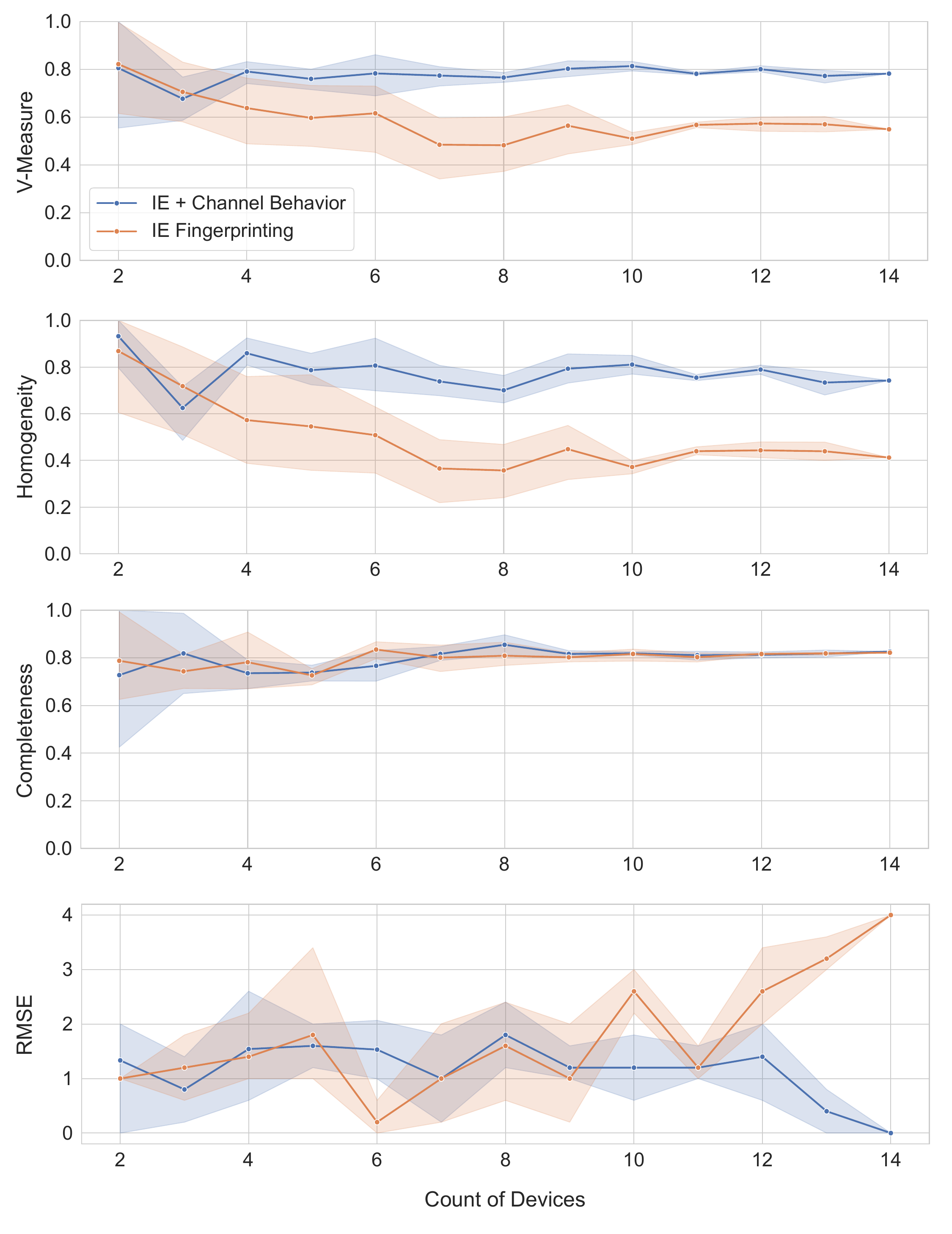}
    \caption{Comparing Clustering Results: Twin Companion dataset}
    \label{fig:clustering-metrics-identical}
\end{figure}

\begin{figure}[t]
    \centering
    \includegraphics[width=1\linewidth]{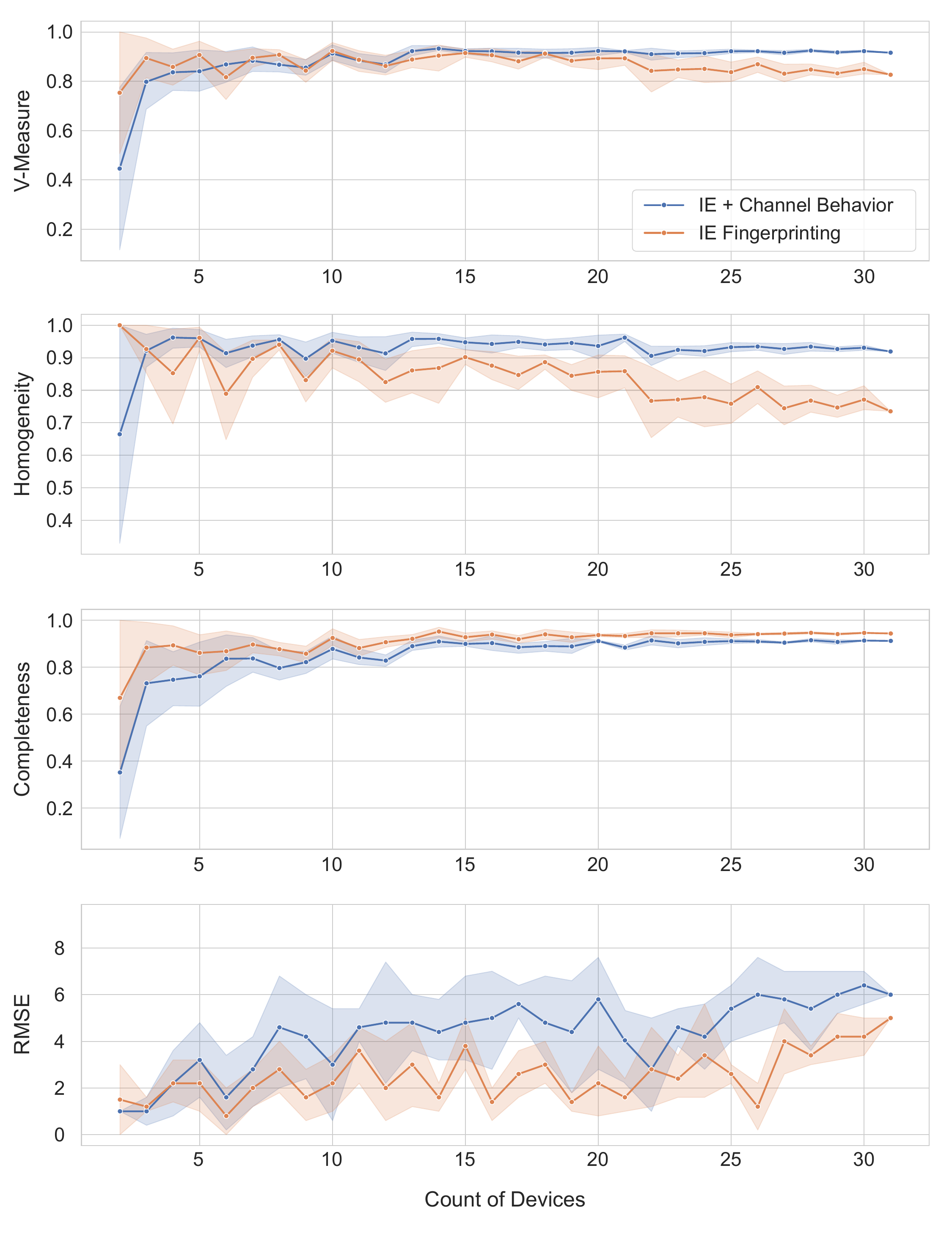}
    \caption{Comparing Clustering Results: Complete dataset.}
    \label{fig:clustering-metrics-complete}
\end{figure}

\section{Experimental results}
\label{sec:results}
Our analysis has two main goals. Our primary objective is to investigate the potential of channel-related behaviors in identifying individual devices. We want to determine whether channel-utilization patterns can be exploited to differentiate devices effectively. In parallel, our secondary goal involves assessing the performance of state-of-the-art methods in the context of identical devices. By doing so, we aim to provide valuable insights into the comparative effectiveness of existing techniques when dealing with devices that share the same model and OS. This analysis contributes to the broader understanding of Wi-Fi device tracking and identification in real-life scenarios. 

\subsection{Testing methodology}
The \textit{Twin Companions} and the \textit{Complete} datasets contain 14 and 31 devices, respectively. To test the performance of the proposed methods at different sizes of the population of devices, we adopt the following strategy: for each target population size $p = 1..P-1$, where $P$ is the maximum number of devices in the original dataset, we produce $d=10$ different subsets by  selecting $p$ devices at random. Then we run the proposed two-stage clustering methodology over each subset, keeping track of the final homogeneity, completeness, and V-measure as well as the Root Mean Squared Error (RMSE) between the final number of created clusters and $p$. For comparison with the state-of-the-art, we also run the algorithm presented in \cite{pintor2022analysis} which uses only IEs features and DBSCAN as the clustering algorithm. In this case, we use the strategy adopted in the proposed method for hyperparameter tuning.
For each value of $p$ we compute the average and standard deviation of the performance metrics and we show them in Figure \ref{fig:clustering-metrics-identical} and \ref{fig:clustering-metrics-complete} for the \textit{Twin Companions} and the \textit{Complete} datasets, respectively.

\subsection{Clustering results}
\subsubsection{Twin companions dataset}
We first comment on the results in Figure \ref{fig:clustering-metrics-identical}, obtained on the more challenging dataset containing many identical devices. As one can see, the use of the proposed method allows us to outperform traditional approaches based only on the clustering of IEs features. Indeed, the use of multi-channel features to differentiate between devices of the same type allows for increased cluster homogeneity, and in turn the V-measure, at all tested population sizes. The proposed method also surpasses IEs fingerprinting in terms of RMSE. Therefore, it finds use in accurate device counting in the presence of MAC randomization.   

\subsubsection{Complete dataset}
Figure \ref{fig:clustering-metrics-complete} illustrates the results for the complete dataset compared to the same IEs fingerprinting technique. Also in this case the proposed method outperforms IEs fingerprinting in terms of cluster homogeneity and V-measure at all population sizes $p$. The difference is particularly highlighted at large population sizes when the likelihood of having identical devices is greater. For what concerns the RMSE, the proposed technique is characterized by a slightly higher error compared to IEs fingerprinting, generally overestimating the actual population by two more devices. 

\subsection{Discussion and limitations}
While IEs fingerprinting proves effective in distinguishing between devices of different models, our analysis reveals a limitation in its ability to differentiate between identical device models. In both the scenarios of the identical devices dataset and the complete dataset, we observe that devices with the same model are grouped within the same cluster. This grouping results in an estimated number of devices that is consistently less than the actual count. Our proposed methodology improves clustering performance and reduces RMSE in estimating device numbers for the identical dataset. At the same time, our methodology presents two main limitations. The first is related to the utilization of cosine similarity as a distance metric for the feature vectors $F$. Generally, longer vectors yield more precise device signatures. However, we observed a specific behavior in certain iPhones running iOS \texttt{16.4.1(a)}. These devices generate shorter Probe Request bursts, often consisting of only 2-4 different frames, primarily on the same channel. Consequently, when calculating the cosine similarity for very short arrays padded with numerous zeros, effectively distinguishing probes from the same device becomes challenging due to their high similarity.

The second limitation arises from devices that present multiple, distinct, yet consistent behaviors. In our analysis, these devices are often counted multiple times, with each unique behavior considered as a separate device. This practice may lead to overestimating the total number of devices, particularly in cases where devices display diverse patterns.
\section{Conclusion}
\label{sec:conclusion}
This paper significantly expanded an open dataset of Wi-Fi Probe Requests, mainly focusing on devices of the same model. This expanded dataset provides a valuable resource for further research in device identification and clustering. We publicly release the dataset for reproducible research\footnote{https://github.com/GiovanniBaccichet/ProbingPatternsDataset}.
The research presents a new technique for Wi-Fi Probe Request fingerprinting that makes use of multi-channel behavior patterns and IE fingerprinting, which is a feature that has yet to be widely considered in the state-of-the-art. This method aims to improve the accuracy of device identification, especially for devices with the same model. The study conducted a comprehensive comparison of the proposed technique with existing state-of-the-art methods. It demonstrated that the novel approach offered enhanced clustering performance and reduced RMSE in estimating the number of devices, even when dealing with identical device models. As future works, we plan to (i) continue increasing the dataset size by adding other devices and (ii) target some of the limitations of our approach by introducing physical layer features (RSSI, CSI).

\section*{Acknowledgement}
This study was carried out within the MICS (Made in Italy
Circular and Sustainable) Extended Partnership and received
funding from Next-Generation EU (Italian PNRR M4 C2,
Invest 1.3 – D.D. 1551.11-10-2022, PE00000004). CUP MICS
D43C22003120001. Additional fundings received by PRIN
project COMPACT, CUP: D53D23001340006.

\bibliographystyle{IEEEtran}
\bibliography{IEEEabrv.bib, bibliography.bib}

\end{document}